\documentclass[aip,pop, preprint, superscriptaddress]{revtex4-1}

\usepackage{amsmath}
\usepackage{graphicx}
\usepackage{dcolumn}

\begin{document}

\title{Enhancement on Laser Intensity and Proton Acceleration Using Micro-tube Plasma Lens Targets}

\author{J. Snyder}
\email[]{snyder.1106@osu.edu}
\author{L. L. Ji}
\author{K. U. Akli}

\affiliation{Physics Department, The Ohio State University, Columbus, OH}

\date{\today}

\begin{abstract}
A hollow cylindrical micron-scale structure is proposed to enhance and manipulate the laser plasma interaction. It is shown through 3-D particle-in-cell simulations that the incident laser pulse intensity is enhanced within the tube. A detailed study of the intensification optimizes the tube dimensions and provides a characterization of the in-tube intensity. By coupling the micro-tube plasma lens to a traditional flat interface, we show an increase in on-target intensity. We detail proton energy enhancement as a potential application of the micro-tube plasma lens target, where the tube structure acts to focus the light and provide additional electrons that enhance the accelerating sheath field.
\end{abstract}


\maketitle

\section{Introduction} The interaction of intense laser pulses with structured interfaces is of great interest due to its applications in the generation of high energy electrons, protons, and x-rays. Experimentally, structured target interaction with high intensity laser pulses showed an enhancement in laser absorption \cite{kulcsar2000intense,purvis2013relativistic,ivanov2015enhanced}. Recently, snow targets have shown the ability to enhance proton energies when compared to flat interfaces\cite{zigler2013enhanced}. Silicon micro-wire array towers have been used to demonstrate an enhancement in electron energy and high energy electron yield \cite{jiang2015microengineering}. Cone targets have long been studied for their effects on electron transport \cite{sentoku2004laser,hu2015enhanced,higginson2015high}.

As technology that can be used for target fabrication has progressed\cite{fischer2013three}, advanced structured interfaces have become an active area of theoretical and simulated research. The use of cone targets and cone-tube targets has shown an enhancement in energy and directionality of proton beams\cite{xiao2016enhanced,yang2014generation,kluge2012high,gaillard2011increased,wu2013effect}. Front surface structures, such as towers and near-critical density plasmas, have shown an enhancement in proton energy as well \cite{bin2015ion,brantov2013laser,wang2009high,nodera2008improvement}. Micro-wire array targets were suggested to create a highly directed energetic electron beam with enhanced bremsstrahlung radiation \cite{jiang2014effects,jiang2014enhancing} . The use of hollow cylindrical targets has been demonstrated to produce bright x-rays and attosecond pulses\cite{yi2015bright,lecz2015attospiral}. Additionally, rear-surface structures have been demonstrated to collimate secondary electron and ion beams\cite{wang2014collimated,li2015transport,zou2015control}. By making use of ultrahigh contrast lasers made possible by XPW \cite{jullien2006highly} or plasma mirrors\cite{dromey2004plasma,thaury2007plasma}, it is possible for structured interfaces to effectively manipulate the laser-plasma interaction in ways that remain unexplored and underutilized.

In this article, we present a characterization of the laser pulse inside a micro-tube plasma (MTP) target. The MTP acts to redistribute the incident pulse intensity within the tube, resulting in a localized intensification of the incident pulse. We explore the effects of varying the MTP diameter, which allow us to determine the distance from the entrance of the tube to the intensified hot spot. We characterize the intensified in-tube laser pulse through particle-in-cell (PIC) simulations. Then, by placing a foil at the rear of the optimized MTP lens, the on-target intensity can be greatly enhanced. Enhanced proton energy is studied as a potential application of the MTP lens, where we show an increase in maximum attainable proton energies up to a factor of 3.4 in the optimized case. 

\section{Simulation Setup} 
\begin{figure}[h!]
	\centering
	\includegraphics[width=16cm]{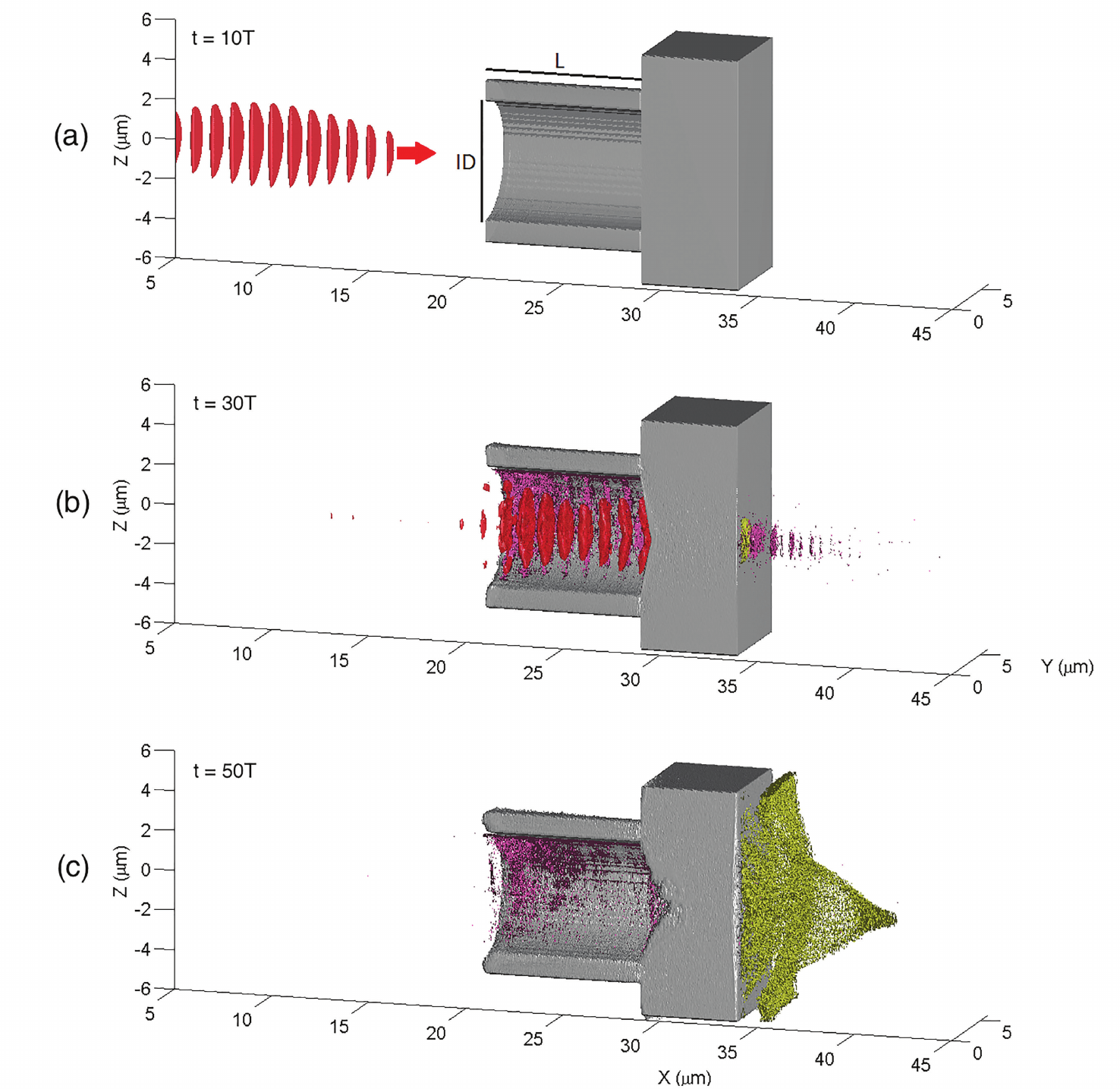}
	\caption{Target conditions for the simulation. Distributions of target electron density (gray), in-tube electron density (purple), proton density (green), and laser intensity (red) at t=10T (a), 30T (b), and 50T (c), respectively. The laser enters from the left of the simulation and is incident on a micro-tube plasma (MTP) lens coupled to a 5$\lambda$ thick CH$_2$ foil. The length (L) and the inner diameter (ID) of the MTP lens are varied to locate the intensity spike on the rear substrate and optimize the on-target intensity. \label{fig:isosurface}}
 \end{figure}

Simulations were performed with the 3D PIC code VLPL\cite{pukhov1999three}. For ion acceleration simulations, a y-polarized laser pulse propagates along the x-direction in a simulation box of 75$\lambda$ $\times$ 12$\lambda$ $\times$ 12$\lambda$ in x $\times$ y $\times$ z, where $\lambda = 0.8\mu m$ is the laser wavelength. During MTP optimization simulations, the x-dimension of the simulation box was varied to decrease simulation time, while the y- and z-dimensions remained constant. The cell size was held constant at 0.02 $\times$ 0.1 $\times$ 0.1 in the x $\times$ y $\times$ z dimensions for all simulations. A time step of $\Delta t = 0.008T$, where $T=2\pi / \omega_0$ ($\omega_0$ is the laser frequency) is chosen to meet the resolution criterion for relativistic electron motion \cite{arefiev2015temporal}. A transversely super-gaussian laser pulse with pulse profile $a_y = a_0 e^{-(r/\sigma)^4 - (t/\tau)^2}$ is focussed to a position of 29$\lambda$ onto a CH$_2$ substrate foil with thickness $5\lambda$. The foil thickness is chosen such that for all geometries studied, the dominant ion acceleration mechanism is expected to be target normal sheath acceleration (TNSA). The laser amplitude is initially $a_0 = 50$ where $a = eE_l/m_e\omega_0c$ , duration $\tau = 12T$, and focal spot $\sigma = 3\lambda$, which are chosen to closely match the parameters of the Scarlet laser housed at Ohio State University \cite{poole2014liquid,willis2015confocal}. On the front surface of the foil, we position carbon microtubes of varying length (L) and inner diameter (ID) with a wall thickness of $\lambda$, as shown in Figure \ref{fig:isosurface}a. The length of a given microtube ID was chosen based on optimization of the intensification factor as discussed below. The electron density of the tube is chosen to be $n_e = 180n_c$ ($n_c = m_e \epsilon_0 \omega^2/e^2$ is the critical plasma density) to imitate the electron density of tubes that are easily manufactured using 3-D printing techniques. The rear CH$_2$ foil is initialized to $n_e = 150n_c$. The whole target is initially cold and fully ionized.

\section{Simulation Results} The full detail of light intensification within a MTP lens can be found elsewhere \cite{ji2015towards}. In short, a laser pulse incident on a hollow cylindrical aperture will undergo diffraction. In the near-field Fresnel region, the result is a boost of the intensity within the cylinder resulting from a redistribution of the incident pulse. Previous work showed by varying the pulse intensity while maintaining the tube geometry that there is an intensity dependent intensification factor, where as the incident pulse becomes more intense the intensification factor increases. The intensity dependence results from additional diffraction caused by tube electrons being dragged into the hollow region of the MTP. In order to optimize the intensification effect, we perform simulations on MTP targets with variable dimensions. Since diffraction is dependent on the geometrical properties of the hollow cylinder, we fix the incident pulse parameters while varying the tube dimensions to characterize the in-tube laser profile. Then, using the MTP structure as a focusing element, we place an optimized MTP on the front surface of a flat CH$_2$ foil and investigate the MTP effects on proton acceleration.

\subsection{Tube Length Optimization and Characterization} 
\begin{figure}[h]
	\centering
	\includegraphics[width=16cm]{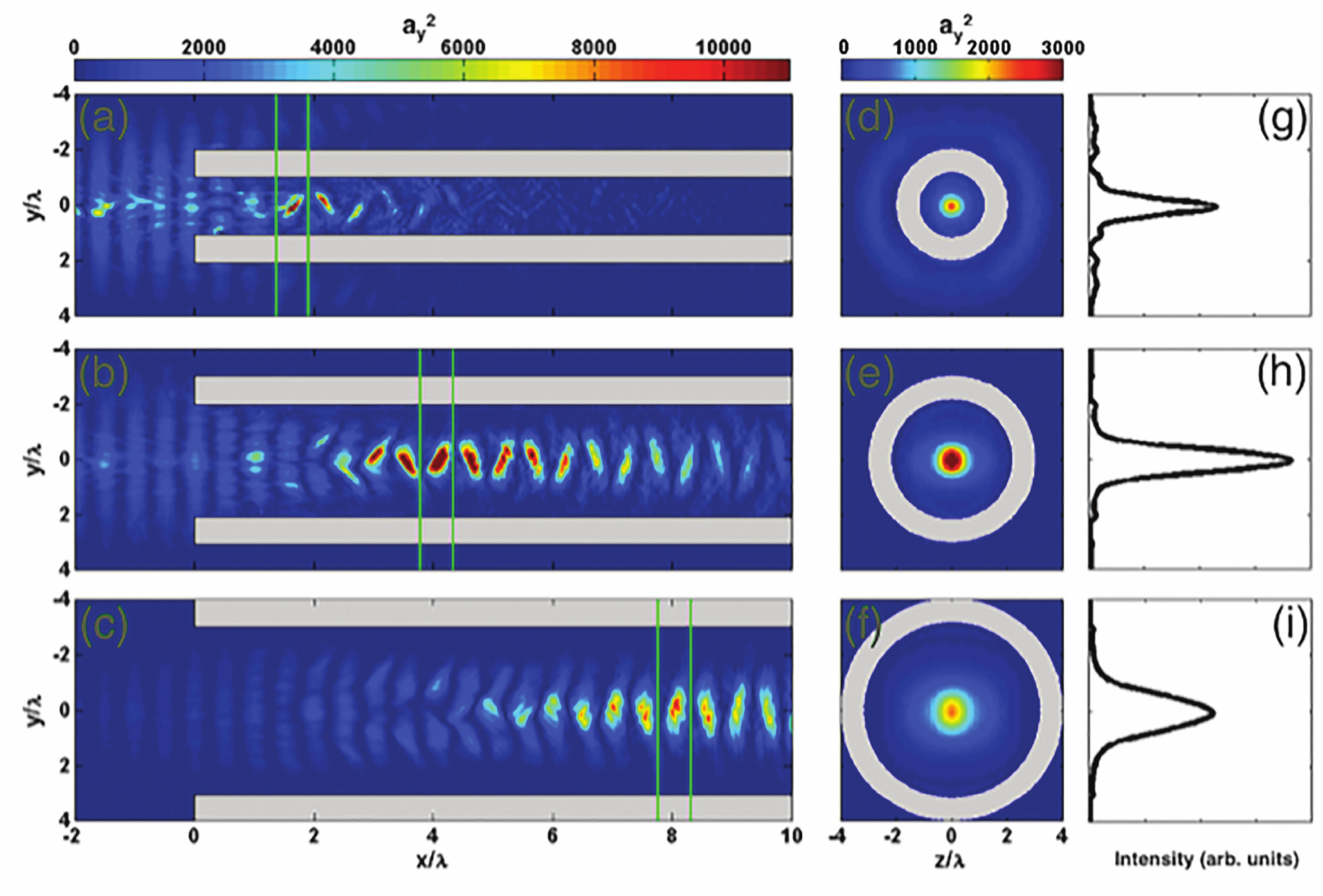}
	\caption{Intensity distribution cutout at z=0 in the x-y plane at the time of peak intensity for the inner diameter of (a) 2$\lambda$, (b) 4$\lambda$, and (c) 6$\lambda$. (d)-(f) shows the transverse intensity distribution for the corresponding tube dimensions given in (a)-(c), averaged in the laser propagation direction within the green box as described in the text. Transverse intensity lineouts for the average intensity is shown in (g)-(i) on an arbitrary scale. \label{fig:intensity_tube}}
 \end{figure}

By varying the ID of the MTP target, the location and level of intensification varies greatly. As such, we conducted an extended study on the optimal relation between tube ID and length. The tube ID was studied without the rear CH$_2$ in order to determine the field as seen at the location of peak intensity, the results of which are summarized in Table \ref{tbl1}. The ID was varied from 2-7$\lambda$ using a laser pulse as described above. The intensification factor, defined as $\eta_{peak} = I_{peak,\, in\, tube}/I_{peak,\, input}$, is noted for each tube diameter studied. With a smaller tube, the diffraction effect coupled with the background plasma focussing effect causes the peak intensity to be closer to the entrance of the tube. As the ID is increased, the focussing position is shifted farther from the entrance of the tube. A peak intensity snapshot is shown in Figure \ref{fig:intensity_tube} (a)-(c). As shown, the highest peak intensity from the geometries  studied is found with the 2$\lambda$ ID tube ($\eta_{peak} =8.56$), although simulation results suggest this intensification is short lived, as detailed later in the text. The 4$\lambda$ ID tube also demonstrates an exemplary peak intensification ($\eta_{peak}=8.36$), with the peak intensity falling lower for the larger ID tubes. In order to fully characterize the in-tube pulse profile, we examine the lifetime of the focussing effect to determine the influence of increase background plasma that is found at smaller tube ID.

In order to characterize the focal spot lifetime, we locate the position and time of peak intensity within the micro-tube. We record the intensity in the entire simulation every half laser period. At the time of the peak intensification, we average the intensity longitudinally in space over  $x=\pm \lambda/4$ from the location of peak intensity, as shown in Figure \ref{fig:intensity_tube} (a)-(c), where the green box indicates the location where the averaging takes place. This is repeated for $\pm 12$ laser periods from the peak time in order to capture the majority of the beam. After this averaging, we note the peak value of the averaged intensity, $I_{ave,\, in \, tube}$. We define the intensification of the beam averaged in this way as $\eta_{ave} = I_{ave, \, in \, tube}/I_{ave, \, input}$, where $I_{ave, \, input}$ is the peak value of the input pulse average in the same manner as $I_{ave, \, in \, tube}$. The results from this averaging technique is shown in Figure \ref{fig:intensity_tube} (d)-(f), while line-outs of these intensity distributions are shown in (g)-(i). We find that $\eta_{ave}$ is best for $ID=4\lambda$ with a value of 5.0, while the average intensification falls off on either side of this peak. The 4$\lambda$ ID matches the peak intensity with hot-spot duration well, resulting in an $\eta_{peak}$ greater then 8 and an $\eta_{ave}$ up to 5.

\begin{table*}[t]
\centering
\setlength{\tabcolsep}{.035\textwidth}
\begin{tabular}{ccccccc}
\hline
Case & ID & Length &  $\eta_{peak}$ & $\eta_{ave}$ & $\sigma$ (FWHM) & E$_{p,max} (MeV)$\\[3pt]
\hline
0 & - & - & - & - & 3.6$\lambda \times$3.6$\lambda$ & 66\\
1 & 2$\lambda$ & 1.6$\lambda$ & 8.56 & 3.20 & 0.7$\lambda \times$0.8$\lambda$ & 104\\
2 & 3$\lambda$ & 2.2$\lambda$ & 7.40 & 3.44 & 0.9$\lambda \times$1.0$\lambda$ & 123 \\
3 & 4$\lambda$ & 4.0$\lambda$ & 8.36 & 5.00 & 0.9$\lambda \times$1.0$\lambda$ & 167 \\
4 & 5$\lambda$ & 5.1$\lambda$ & 5.68 & 3.44 & 1.3$\lambda \times$1.4$\lambda$ & 180 \\
5 & 6$\lambda$ & 8.0$\lambda$ & 4.22 & 3.10 & 1.4$\lambda \times$1.5$\lambda$ & 232\\
6 & 7$\lambda$ & 10.5$\lambda$ & 2.41 & 2.15 &  1.7$\lambda \times$1.9$\lambda$ & 228\\
7 & - & - & 3.15 & 3.24 &  1.5$\lambda \times$1.5$\lambda$ & 78\\[3pt]
\hline
\end{tabular}
\caption{Beam characterization within the micro-tube plasma lens. Case 0 details the input pulse, while cases 1-6 show the effects of varying the ID of a micro-tube plasma lens. The length column represents the distance from the entrance of the MTP lens where the peak intensity occurs. The peak intensification is defined as $\eta_{peak} = I_{peak,\, in\, tube}/I_{peak,\, input}$. The average intensification, $\eta_{ave} = I_{ave, \, in \, tube}/I_{ave, \, input}$, is calculated by an averaging technique described in the body of the paper. The spot size is listed as the FWHM in $y\times z$ for the pulse averaged in space over a half wavelength at the time of peak intensity. Case 7 details the characterized pulse with $a_0=87$ used to determine the effects of light intensification without the MTP plasma effects included.\label{tbl1}}
\end{table*}
 
\subsection{Proton Acceleration Enhancement}
\begin{figure}
	\centering
	\includegraphics[width=8cm]{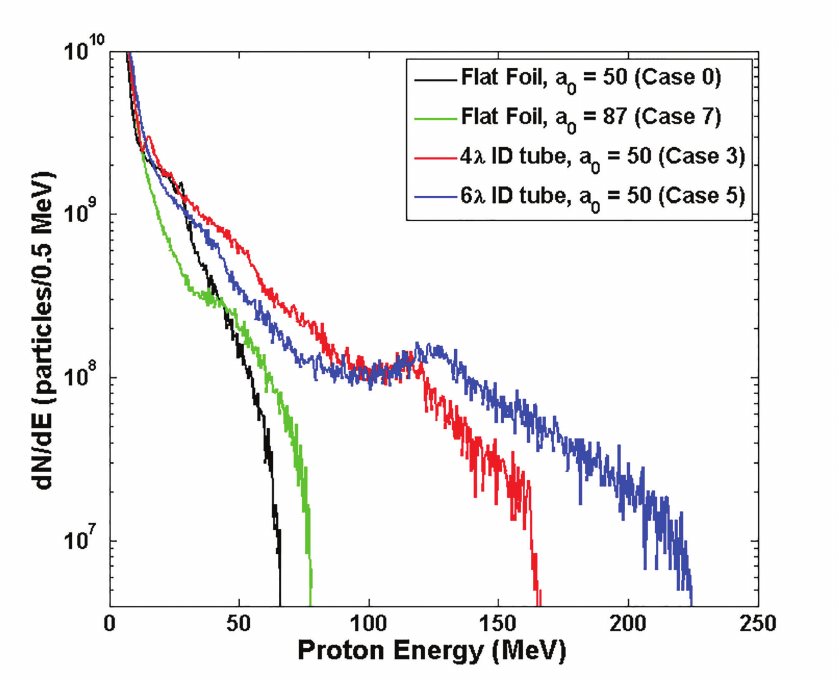}
	\caption{Proton energy distribution for a flat interface with $a_0 =50$ (case 0, black), a flat interface with $a_0 =87$ (case 7, green), a 4$\lambda$ ID MTP lens target (case 3, red), and a 6$\lambda$ ID MTP lens target (case 5, blue) at T=90. The most energetic ions are found using the optimized parameters for a 6$\lambda$ ID MTP target. \label{fig:proton_en}}
 \end{figure}

With a well characterized interaction between the incident pulse and variable MTP lenses, a series of simulations coupling a CH$_2$ foil to the location of the peak intensity have been performed. The proton energy spectrum for four representative cases is shown in Figure \ref{fig:proton_en}  at $t=90T$, where the incident pulse is focussed on the front of the CH$_2$ foil at $t=30T$. For reasons shown below, although the optimized 4$\lambda$ gives the best on-target intensity, this does not necessarily equate to the best ion energy spectrum with the rear foil parameters studied in this work. Instead, the highest energy protons found in our study arise from the $6\lambda$ ID MTP. This result indicates that the enhancement of TNSA does not solely arise from the intensified on-target laser field, but is more likely a combination of several effects. As is known, TNSA is more relied on hot electrons but less sensitive to the laser intensity ($\sim \sqrt{I}$). It is thus natural to assume that introducing the tube also enhances the generation of energetic electrons.

We identify two main components that are responsible for the enhanced ion energy: a higher on-target intensity and an enhanced sheath field. In an effort to elucidate the effects from the contributing factors, an additional simulation with an altered incident pulse onto a flat CH$_2$ was performed. In order to determine the effects of the high on-target intensity independently of the plasma effects of the tube, we used the beam conditions as characterized at the peak location in the tube to irradiate a flat foil. 

\subsubsection{Increased On-Target Intensity}
In the TNSA model, the maximum ion energy scales as $E_{ion,max}$ $\alpha$ $I^{1/2}$ \cite{hatchett2000electron}. If the enhancement in maximum ion energy were due only to the intensification within the tube, the highest proton energy would arise from the $4\lambda$ ID tube. This is not the case, as the $6\lambda$ ID tube produces a maximum proton energy of 232 MeV, whereas the $4\lambda$ ID MTP target achieves a peak proton energy of 166.5 MeV. In an effort to understand the effect of intensification on the proton energy distribution, we perform a simulation using a laser pulse with $a_0 = 87$ interacting with a flat $CH_2$. As shown in Table \ref{tbl1}, the pulse matches reasonably well with the characterized pulse for our initial investigations with the $6\lambda$ ID MTP lens. The proton energy spectrum from this simulation is shown in green in Figure \ref{fig:proton_en}. The maximum proton energy from the $a_0 = 87$ simulation reaches 78 MeV, which is slightly higher than that of the $a_0 = 50$ simulation (black, 66 MeV), but not nearly as high as the maximum energy achieved with optimized MTP targets. Clearly, the dominant effect of proton enhancement cannot be attributed solely to the intensification in this target regime.
 
\subsubsection{Enhanced Sheath Field}
\begin{figure}[h!]
	\centering
	\includegraphics[width=16cm]{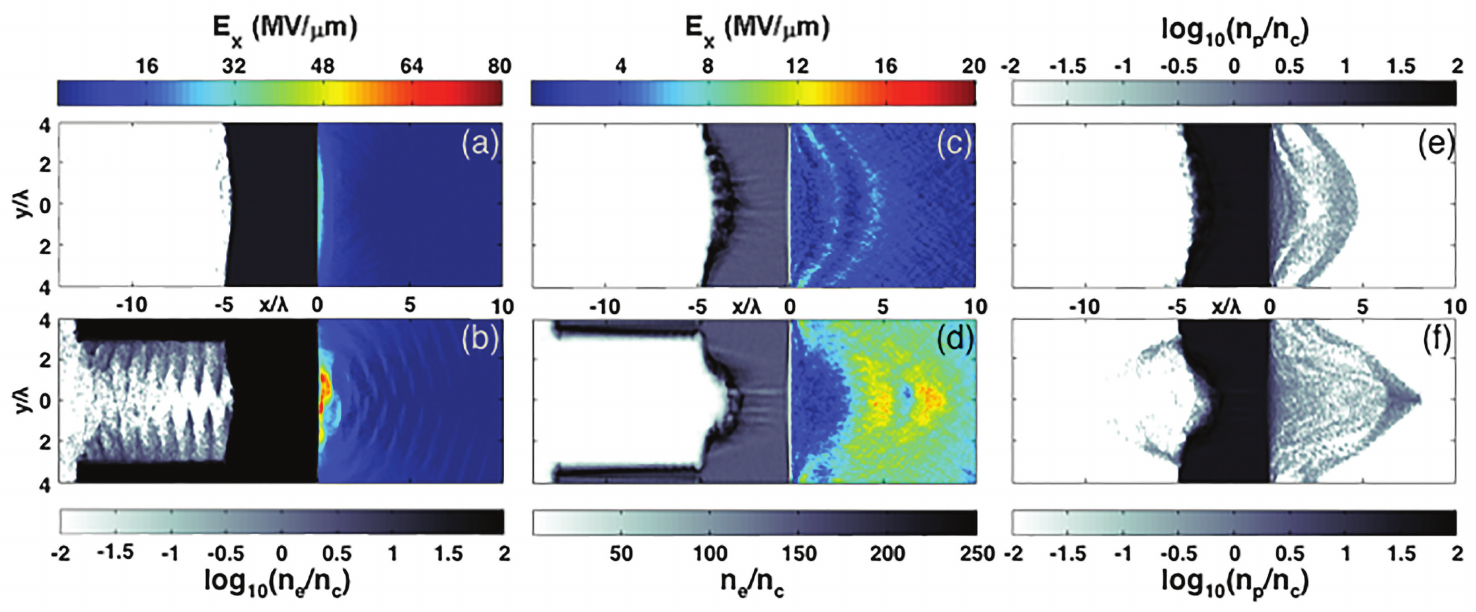}
	\caption{Electron density (grayscale) and the longitudinal electric field E$_x$ (color scale) in the x-y plane at z=0 for a (a),(c) flat foil and (b),(d) 6$\lambda$ ID MTP lens target (Case 5) at the time the peak of the laser pulse reaches the front of the CH$_2$ foil (a)-(b) and 20T later in the simulation (c)-(d). Note the log scale of electron density in (a)-(b) to show the periodic electron bunches located in the tube, while the linear scale of electron density in (c)-(d) highlight the increased front-surface target deformation when using the MTP lens target. The proton distribution on a log scale 20T after the peak reaches the front of the CH$_2$ foil is shown in (e)-(f). \label{fig:compare}}
 \end{figure}

\begin{figure}[h!]
	\centering
	\includegraphics[width=8cm]{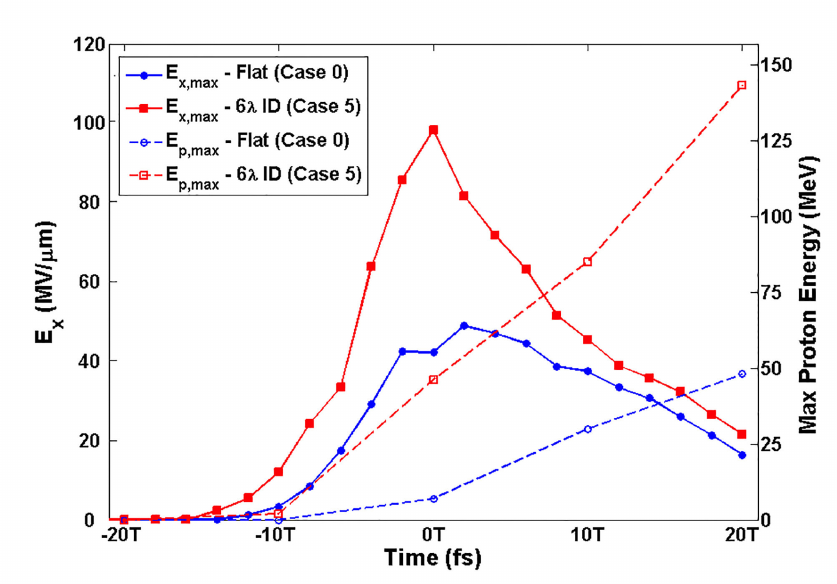}
	\caption{Maximum sheath field (solid line) and proton energy (dashed line) evolution for the flat $CH_2$ foil (Case 0, blue) and the optimized 6$\lambda$ MTP lens target(Case 5, red) where 0T corresponds to the peak of the incident pulse reaching the front surface of the CH$_2$ foil.\label{fig:ex_proton_en}}
 \end{figure}

The substrate thickness dictates that the majority of high energy protons originate from the rear of the $CH_2$ foil. As such, the dominant factor in the proton acceleration is the sheath field. In order to understand the sheath field effect, we look at the electric fields that arise at the rear of the foil in the $6\lambda$ ID MTP target and compare this to the flat CH$_2$ foil. Figure \ref{fig:compare} shows the longitudinal electric fields for the two cases at $t=30T$ and $t=50T$. The $6\lambda$ ID MTP target has a sheath field that not only has a higher peak value ($>$100 MV/$\mu$m) as evident by Figure \ref{fig:compare} (a)-(b), but also has an extended longitudinal range as seen in the (c)-(d) of the same figure.
 Additionally, we compare the maximum sheath field value for the two cases every two laser periods through the interaction of the laser pulse (Figure \ref{fig:ex_proton_en}). The sheath field in the $6\lambda$ ID case develops at an earlier time than the flat $CH_2$, and reaches a peak value that is nearly double that of the flat foil. It is worth noting that while the increased on-target intensity is not the primary contributing factor to enhanced TNSA, the strong deformation of the substrate interface in the MTP case in Figure \ref{fig:compare} (d) indicates an increased hole-boring velocity compared to the flat interface in (c). This is a direct consequence of the enhanced on-target intensity with the MTP lens target.
  
The sheath field enhancement is attributed to high density electron bunches that are pulled out of the tube from the transverse laser electric field and accelerated forward via direct laser acceleration (DLA) of the laser pulse\cite{kluge2012high}. When the laser pulse is sufficiently intense, these electrons can stay in the acceleration phase for a longer time. As the laser reaches the critical surface of the rear foil, the electrons decouple from the laser, and move through the rear foil, as seen in Figure \ref{fig:isosurface}b.  These electrons are clearly visible in Figure \ref{fig:compare}b, and act to periodically enhance the sheath field as shown in the colorscale of the Figure \ref{fig:compare}b.  The high density, highly localized electrons result in an electric sheath field that is far greater than one would expect from bulk hot electrons typically associated with TNSA of ions as described by the ponderomotive scaling \cite{wilks1992absorption}. Additionally, the sheath field peak value at later times (Figure \ref{fig:compare}(c)-(d)) is similar in the two cases; however, the longitudinal extent of the sheath field is greatly enhanced in the case of the MTP lens target. The simulation results demonstrate that the confinement of localized electron bunches that result from the MTP lens structured interface give rise to enhanced sheath fields and ultimately, much higher proton energies.
 
\section{Conclusion}
In conclusion, we investigated the interaction of a highly relativistic laser pulse with micro-tube plasma lens coupled to traditional flat interfaces. By varying the dimensions of the MTP lens, we characterize the in-tube laser pulse to optimize the on-target intensity. Finally, we establish a potential application of the MTP lens target by demonstrating an enhancement in proton energy when comparing to traditional flat interfaces. The enhanced ion acceleration is a results from laser pulse intensification as well as localized electron bunches that are guided by the MTP walls. Other intensity favorable mechanisms such as radiation pressure acceleration (RPA)\cite{robinson2008radiation,macchi2009light}, break-out afterburner acceleration (BOA)\cite{yin2006gev}, and X-ray generation\cite{ji2014energy} would likely benefit more so than TNSA from the MTP target, and these studies are currently underway. Our results show that with current laser and 3D printing technology, it is possible to increase the on-target intensity and efficiently acceleration ions to significantly higher energies.

\section{Acknowledgements}
This work was supported by the AFOSR (Air Force Office of Scientific Research) under contract No. FA9550-14-1-0085. The authors would like to thank Alexander Pukhov for the use of the code VLPL. 

\bibliography{bib1}
\bibliographystyle{apsrev4-1}
\end{document}